\documentclass{ws-procs975x65}            
\begin{document}
\title{
Paradoxes of cosmological physics \\
in the beginning of the 21-st century}

\author{Yurij Baryshev$^*$ }

\address{Astronomical Institute,
Saint Petersburg State University,\\
Saint Petersburg, 198504, Russia\\
$^*$E-mail: yubaryshev@mail.ru}

%

\begin{abstract}

In the history of cosmology physical paradoxes played important role for development of contemporary world models. Within the modern standard cosmological model there are both observational and conceptual cosmological paradoxes which stimulate to search their solution. Confrontation of theoretical predictions of the standard cosmological model with the latest astrophysical observational data is considered.  A review of conceptual problems of the Friedmann space expending models, which are in the bases of modern cosmological model, is discussed. The main paradoxes, which are discussed in modern literature, are the Newtonian character of the exact Friedmann equation, the violation of the energy conservation within any comoving local volume, violation of the limiting recession velocity of galaxies for the observed high redshift objects. Possible observational tests of the nature of the cosmological redshift are discussed.

\end{abstract}


\bodymatter

\section{The standard cosmological model}\label{aba:sec1}
Nowadays the expanding Big Bang cosmological model is generally accepted as
the standard cosmological model (SCM) for description of the structure and evolution of the physical Universe
(Peebles\cite{peebles93}, Weinberg\cite{weinberg08},
Baryshev \& Teerikorpi\cite{bartee12}).
SCM is based on the geometrical gravity theory (general relativity) and uses
the description of all physical processes in expanding space.
The fundamental assumptions of the SCM are:

$\bullet$ Homogeneous and isotropic matter distribution in the
expanding Universe

$\,\,\,$ ($\rho=\rho(t)$;  $p=p(t)$; $g^{ik} = g^{ik}(t)$)
\footnote{We use main definitions and notations similar to
Landau \& Lifshitz\cite{ll71},
so 4-dimensional tensor indices are denoted by Latin letters
$i,k,l...$ which
take on the values 0, 1, 2, 3, and the metric has signature
$(+,-,-,-)$.}
.

$\bullet$  General relativity is applicable to the whole Universe
($g_{ik}$; $\:\Re_{iklm}$; $\:T^{ik}_{(\mathrm{m+de})}$).

$\bullet$  Laboratory physics can be extended into the expanding space.

$\bullet$  Inflation in the early Universe is needed for flatness, isotropy and for initial

$\,\,\,$ conditions of large scale structure formation.

\subsection{Einstein's cosmological principle }
The fundamental basic element of the SCM is the Einstein's Cosmological Principle, which states that \textbf{the universe is spatially homogeneous and isotropic at enough  "large scales"}. The term "large scales" relates to the fact that the universe is obviously  inhomogeneous at scales of galaxies and clusters of galaxies. The hypothesis of homogeneity and isotropy of the matter distribution in space means that starting from certain scale $r_{\mathrm{hom}}$, for all scales
$r>r_{\mathrm{hom}}$ we can consider the total energy density
$\varepsilon=\rho c^2$  and the total pressure $p$ as a function of time only, i.e. $\varepsilon(r,t) = \varepsilon(t)$ and $p(r,t) = p(t)$ . Here the total energy density and the total pressure are the sum of the energy densities for matter and dark energy :
$\varepsilon = \varepsilon_{\mathrm{m}} + \varepsilon_{\mathrm{de}}$, and
$p = p_{\mathrm{m}} + p_{\mathrm{de}}$.

An ideal fluid equation of state
$p=\gamma \varrho c^2$
is usually considered for cosmological fluid,
where usual matter and dark energy have following partial equations of state:
$p_{\mathrm{m}} = \beta \varepsilon_{\mathrm{m}}$ with  $0\leq \beta \leq 1$, and
$p_{\mathrm{de}} = w \varepsilon_{\mathrm{de}}$ with  $-1 \leq w < 0$.
Recently values $w<-1$ also were considered for description the "fantom" energy.

\subsection{Expanding space paradigm}
An important consequence of homogeneity and isotropy is that the line element
$ ds^2 = g_{ik} dx^i dx^k$
may be presented in the Robertson-Walker form:
\begin{equation}
 ds^2 = c^2dt^2 - S^2(t)d\chi^2 - S^2(t) I_k^2 (\chi) (d\theta^2 +
 sin^2 \theta d\phi^2)\,,
\label{rw1}
\end{equation}
where $ \chi,\,\theta,\,\phi $ are the "spherical" comoving space coordinates,
$t$ is synchronous time coordinate, and
$I_k(\chi) = (sin(\chi),~\chi,~sinh(\chi))$,
corresponding to curvature constant values $k = (+1,~0,~-1)$ respectively.
$S(t)$ is the scale factor, which determines the time dependence of the metric.

The \emph{expanding space} paradigm states that the proper (internal) metric distance $r$ to a galaxy with fixed co-moving coordinate $\chi$ from the observer is given by relation $r(t) = S(t)\cdot \chi $
and increases with time $t$ as the scale factor $S(t)$.

Note that physical dimension of metric distance [r] = cm , hence, if physical dimension [S] = cm, then $\chi$ is the dimensionless comoving coordinate distance. In direct mathematical sense $\chi$ is the spherical angle and S(t) is the radius of the sphere (or pseudosphere) embedded in the 4-dimensional Euclidean space.
It means that the "cm" (the measuring rod) itself is defined as unchangeable unit of length in the embedding 4-d Euclidean space.

It is important to point out that the hypothesis of homogeneity and isotropy of space implies that for a given galaxy the recession velocity is proportional to distance (\textbf{exact linear velocity-distance relation} for all
RW metrics Eq. (\ref{rw1})):
\begin{equation}
\label{expvel}
 V_{exp}(r) = \frac{dr}{dt} =  \frac{dS}{dt}
 \chi = \frac{dS}{dt}\cdot \frac{r}{S} = H(t) r = c \frac{r}{r_{H}}
\end{equation}
where $H=\dot{S}/S$ is the Hubble constant (also is a function of time) and
$\,r_{H} = c/H(t)$ is the Hubble distance at the time $t$.

\subsection{Geometrical gravity theory }
The Einstein-Hilbert field equations of the general relativity have the form:
\begin{equation}
  \Re^{ik} - \frac{1}{2}\,g^{ik}\,\Re =
  \frac{8\,\pi\,G}{c^4}\,\,(T^{ik}_{(\mathrm{m})} + T^{ik}_{(\mathrm{de})})\,,
\label{efeq1}
\end{equation}
where $\:\Re^{ik}$ is the Ricci tensor, $\:T^{ik}_{(\mathrm{m})}$
is the energy-momentum tensor (EMT) of the matter, which includes
all kinds of material substances, such as particles, fields,
radiation, and $T^{ik}_{(\mathrm{de})}$ is the EMT of dark energy,
in particular, the cosmological vacuum is described by
$T^{ik}_{(\mathrm{vac})} = g^{ik}\Lambda$, where
$\Lambda$ is Einstein's cosmological constant.
Usually $\:T^{ik}_{(\mathrm{m})}$ and $T^{ik}_{(\mathrm{de})}$ are
considered as independent quantities, though there are models with
interacting matter and dark energy\cite{gromov04}.

It is important to note that $\:T^{ik}_{(\mathrm{m})}$
does not contain the energy-momentum tensor of the gravity field itself,
because gravitation in general relativity is a property of space and
is not a material field. This is why there is no such concepts as gravity force
and energy of gravitational field in general relativity.

A mathematical consequence of the field equations (Eq. (\ref{efeq1})) is
that the covariant derivative of the left side equals zero
(due to Bianchi identity), so for the right side we also have
\begin{equation}
  (T^{ik}_{(\mathrm{m})} + T^{ik}_{(\mathrm{de})} )_{\; ;~i} = 0\,.
\label{bianki}
\end{equation}
The continuity equation (Eq. (\ref{bianki})) also gives the equations
of motion for the considered matter.\footnote{
As it was emphasized by Landau \& Lifshitz\cite{ll71} the Eq.(\ref{bianki})
is not a conservation law
of the energy-momentum of the particles plus gravity system, because
$\:T^{ik}_{(\mathrm{m})}$
does not contain the energy-momentum of the gravity field itself.
}

\subsection{Friedmann's equations}
In comoving coordinates the total EMT has the form
$ T_{k}^{i} = diag(\varepsilon,-p,-p,-p)$
and for the case of unbounded homogeneous matter distribution given by metric
Eq. (\ref{rw1}), the Einstein's equations (Eq. (\ref{efeq1})) are directly
reduced to the Friedmann's equations (FLRW – model). From the initial set of 16 equations we have only two independent equations for the (0,0) and (1,1) components, to which we must add the continuity equation (Eq. (\ref{bianki})) which has the form $3\dot{S}/S=-\dot{\varepsilon}/(\varepsilon + p)$.

Using the definition of the Hubble constant $H=\dot{S}/S$ , the Friedmann's equations get the form:
\begin{equation}
\label{friedmann00}
 H^2 - \frac{8\pi
 G}{3}\varrho=-\frac{kc^2}{S^2}\,,\,\,\,\,\,\,\mathrm{or}\,\,\, \,\,\,
  1 - \Omega = -\Omega_k\,,
 \end{equation}
and
\begin{equation} \label{friedmann11}
 \ddot{S}= - \frac{4 \pi G}{3} \left(
 \varrho+ \frac{3p}{c^{2}}\right)S\,,\,\,\,\,\,\,\mathrm{or} \,\,\,\,\,\,
q = \frac{1}{2}\Omega \left(1 + \frac{3p}{\varrho c^2} \right)\,,
\end{equation}
where $\Omega = \varrho / \varrho_{crit}$,
$\varrho_{crit}=3H^2/8\pi G$, $\Omega_k = kc^2/S^2H^2$ è $ q = -
\ddot{S}S/\dot{S}^2 $, and
$\Omega, p ,\varrho$
are the total quantities, i.e. the sum of corresponding components for matter and dark energy.
Solving the Friedmann's equations one finds the dependence on time the scale factor $S(t)$ or the metric distance $r(t)$, which is the mathematical presentation of the space expansion.

\subsection{Fundamental conclusions of the SCM}
There are many explained astrophysical phenomena in the frame of the SCM, such as cosmological redshift of distant objects, cosmic microwave background radiation, large scale structure formation, chemical composition of matter and other.
The main observational conclusions of the SCM are:
\begin{itemlist}
\item   Cosmological redshift $ (1+z) = \lambda_{0}/(\lambda_{1}) =
S_{0}/S_{1}\,,$ and the linear velocity-distance relation $V_{exp}=H\times r$
is the consequence of the space expansion $r(t)=S(t)\times \chi$
of the  homogeneous Universe.
\item Cosmic microwave background radiation is the result of the
photon gas cooling in the expanding space $T(z) = T_0 (1+z)$.
\item  Small anisotropy $\Delta T/T(\theta)$ of the CMBR is determined by
the initial spectrum of density fluctuations which are the source
of the large scale structure of the Universe.
\item The physics of the expanding Universe is described by the LCDM model which
predicts the following matter budget at present epoch: 70\% of unobservable in lab dark energy, 25\% unknown nonbaryonic cold dark matter and 5\% ordinary
matter . Visible galaxies contribution is less than 0.5\%.
\end{itemlist}

\section{Observational puzzles of the SCM}
In spite of evident successes of the SCM there are also observational
facts which present severe problems for the SCM. We emphasize here several such problems which were discussed recently in the literature.

\begin{enumerate}
  \item \emph{Absurd Universe.} The visible matter of the Universe, the part which we  can actually observe, is a surprisingly small (about 0.5\%) piece of the predicted matter content and this looks like an "Absurd Universe"\cite{turner03}. What is more, about 95\% of the cosmological matter density, which determine the dynamics of the whole Universe has unknown physical nature. Turner\cite{turner02} emphasized that modern SCM predicts with high precision the values for dark energy and nonbaryonic cold dark matter, but
      \textit{"we have to make sense to all this"}.
  \item \emph{The cosmological constant problem.}  One of the most serious problem of the LCDM model is that the observed value of the cosmological constant $\Lambda$ is about 120 orders of magnitude smaller than the expectation from the physical vacuum (as discussed by Weinberg\cite{weinberg89} and Clifton et al.\cite{clifton12}). In fact the critical density of the $\Omega =1$ universe is $\varrho_{\mathrm{crit}} = 0.853 \times 10^{-29}$g/cm$^3$, while the Planck vacuum has $\varrho_{\mathrm{vac}} \approx 10^{+94}$g/cm$^3$.

  \item \emph{The cold dark matter crisis on galactic and subgalactic scales.} There are number of problems with predicting behavior of baryonic and nonbaryonic matter within galaxies. It was discussed by Tasitsiomi\cite{tasitsiomi03} and Kroupa\cite{kroupa12} that there are discrepancies between observed and predicted galaxy density profiles (the cusp problem), small number of observed satellites galaxies  (missing satellites problem), and observed tight correlation between dark matter and baryons in galaxies, which is not expected within LCDM galaxy formation theory.
  \item \emph{The LCDM  crisis at super-large scales.} The most recent observational facts which contradict the LCDM picture of the large scale structure formation, come from:  the SDSS and 2dF galaxy redshift surveys (Sylos Labini\cite{sylos11}), problems with observations of baryon acoustic oscillations (Sylos Labini et al.\cite{sylos-bao09}), existence of structures with sizes $\sim$ 400 Mpc/h in the local Universe (Gott et al.\cite{gott05}, Tully et al.\cite{tully14})and
      $\sim$ 1000 Mpc/h radial structures in the very deep galaxy surveys
      (Nabokov \& Baryshev\cite{nabokov10}), existence of qso structures with sizes $\sim$ 500 Mpc/h (Clowes et al.\cite{clowes13}, Einasto et al.\cite{einasto14}), alternative interpretation of the shape of the CMBR fluctuations correlation function
      (Lopez-Corredoira \& Gabrielli\cite{lopez-gab13}),
      lack of CMBR power at angular scales larger 60 degrees and correlation of CMBR quadrupole and octopole with ecliptic plain ( Copi et al.\cite{copi10}), see also recent review by Perivolaropoulos\cite{perivol14}.
\end{enumerate}

\section{Conceptual paradoxes of the SCM}
The existence of the mentioned above observational puzzles in the SCM interpretations of the astrophysical data rises a question:
Does the contemporary standard cosmological model present the ultimate physical picture of the Universe?

Philosophical, methodological and sociological aspects of the development of the science on the whole Universe was recently analyzed by
Lopez-Corredoira\cite{lopez14}, who emphasized the important role of alternative ideas in cosmology, though usually they have small funding in modern scientific society. The mathematical and physical basis for the SCM and alternatives was considered by Baryshev \& Teerikorpi\cite{bartee12}.

As it is natural for progress in physics we should carefully analyze
the fundamental assumptions laying in the basis of the physical theories.
In the Sec.1 we have formulated several fundamental assumptions in the basis of SCM which have led us to the serious observational puzzles (Sec.2).
As was emphasized by Turner\cite{turner02} for making new cosmology one has to answer a new set of questions and the future world model will reveal deep connection between fundamental physics and cosmology: \textit{"There may even be some big surprises: time variation of the constants or a new theory of gravity that eliminates the need for dark matter and dark energy"}\cite{turner02}.

Intriguingly, besides the mentioned above observational puzzles
there are several deep conceptual problems in the foundation of the SCM. Their solution could open the door  to construction more firmly established future cosmology. Below we present several such conceptual difficulties/paradoxes of the SCM, which already have been discussed in the literature:
\begin{itemlist}
\item \textbf{Gravitation field energy paradox:} in the framework of the Einstein's geometrical gravity theory (General Relativity)  there is no physical concept of the energy-momentum density of the gravitational field (also there is no physical concept of the energy quanta of the gravitational field),  though field energy exists for all other fundamental physical interactions.
\item \textbf{1st Harrison's  paradox:} physics of space expansion contains such puzzling phenomena as continuous creation of vacuum, violation of energy conservation, violation of limiting velocity by receding galaxies.
\item \textbf{2nd Harrison's  paradox:} the cosmological redshift in expanding space is not the Doppler effect, but it is a new physical phenomenon which does not tested in the lab, the global gravitational cosmological redshift should be taken into account.
\item \textbf{Hubble-de Vaucouleurs' paradox:} in the expanding space the linear Hubble law is the fundamental consequence of the homogeneity, however modern observations reveal existence of strongly inhomogeneous fractal large-scale galaxy distribution at scales at least  up to 100 Mpc, while the linear Hubble law starts from 1 Mpc, i.e. just inside inhomogeneous structure.
\end{itemlist}

\subsection{Gravitation theory}
Though Einstein's general relativity solved the old gravitational paradox of the Newtonian gravity theory, the geometrical gravity leads to the new form of the paradox at a deep conceptual level - absence of the energy-momentum tensor of the gravitational field.

According to Landau \& Lifshitz\cite{ll71} (paragraph 101 "The energy-momentum pseudotensor")  the Einstein's field equations (Eq.(\ref{efeq1})) should contains \textit{"the four-momentum of matter plus gravitational field; the latter is not included in the expression for $T^{ik}$}". This is why Landau \& Lifshitz
\cite{ll71} claimed that the continuity equation
(Eq. (\ref{bianki}): $(T^{ik})_{\,\,;} = 0$  )
\textit{" does not generally express any conservation law whatever"}.

The "pseudo-tensor" character of the gravity field in GR has remarkable history (see e.g. Baryshev\cite{bar08a}) and had been discussed from time to time for a century, causing surprises for each new generation of physicists. Rejection of the Minkowski space inevitably leads to deep difficulties with the definition  of the energy-momentum for the gravitational field and its conservation.

However this conceptual problem can be solved within non-metric
Feynman's Field Gravitation approach\cite{feynman95} , which is the  theory of the symmetric second rank tensor field (gravitational potentials) in Minkowski space,
and which unite gravity with the other fundamental forces of nature
(consistent development of the FG see e.g. Sokolov \& Baryshev\cite{socbar80} and Baryshev\cite{bar08a}).
As Feynman\cite{feynman95} emphasized: \textit{"geometrical interpretation is not really necessary or essential for physics"} (p.113).

The Feynman's Field Gravitation (FG) theory contains the concept of gravitational EMT and conservation of the total EMT. The geometrical gravity theory may be considered as an approximation of the relativistic quantum field gravity (like geometrical optics for electrodynamics). There are achievable in near future experiments and astrophysical observations which can distinguish between
GR and FG, like additional scalar gravitational radiation (Baryshev\cite{bar08b}).

\subsection{Physics of space expansion}
Mathematically space expansion is a continuous increasing with time of the distance $r(t)$ between galaxies. It is given by relation $r(t)=S(t)\cdot\chi$
where $S(t)$ is the scale factor from Eq. \ref{rw1}. But what does space expansion mean physically?

Cosmological physics of the expanding space is essentially
different from the lab physics and even contains deep paradoxes which should be studied carefully\cite{bar08c}. Physically expansion of the universe means the continues  creation of space together with physical vacuum.
Real Universe is not homogeneous, it contains atoms, planets, stars, galaxies.
In fact bounded physical objects like particles, atoms, stars and galaxies do not expands. So inside these objects there is no space creation. This is why the creation of space is a new cosmological phenomenon, which has not been tested yet in physical laboratory.

The first puzzling feature of the space expansion physics is that the Friedmann's equations Eq. (\ref{friedmann11}, \ref{friedmann00}) in terms of the metric distance $r(t)=S(t)\cdot\chi$ get the exact Newtonian form:
\begin{equation} \label{newton-friedman}
 \ddot{r}= - \frac{G M_{\mathrm{g}}(r) } {r^2} \,,\,\,\,\,\,\,\,\,\,\mathrm{and} \,\,\,\,\,\,\,\,\,
\frac{V^2_{\mathrm{exp}}}{2} - \frac{GM}{r} = \mathrm{const}\,,
\end{equation}
where  $M_{\mathrm{g}}(r) = - \frac{4 \pi G}{3} \left(
 \varrho+ \frac{3p}{c^{2}}\right) r^3 $ is the gravitating mass of a ball with radius $r(t)$. So according to general relativity the dynamics of the whole universe is determined by the exact Newtonian acceleration and Newtonian kinetic plus potential energy conservation (here velocity of light $c$ does not change the Newtonian character of the equations).

 The second puzzling fact of the space expanding universe is that in the case of the equation of state $p=\gamma \varrho c^2$ the mass (energy) of any local ball is changing with time as:
\begin{equation} \label{gravmass-friedman}
M_{\mathrm{g}}(r) = - \frac{4 \pi G}{3} \left(
 1 + 3\gamma\right)\varrho r^3
 \,\,\,\,\,\,\,\,\, \propto \,\,\,\,\,\,\,\,\,S^{-3\gamma}(t)\,.
\end{equation}
For example for photon gas $\gamma = 1/3$ and the mass-energy of the initially hot radiation is cooling proportional to the scale factor $S(t)$.

In cosmology Eq.(\ref{gravmass-friedman}) gives us a possibility to calculate of how much the energy increases or decreases inside any finite comoving volume but it does not tell us where the energy comes from or where it goes.
As Harrison emphasized: \textit{"The conclusion, whether we like it or not, is obvious: energy in the universe is not conserved"} (Harrison\cite{harrison81}, p.276).

Another puzzling consequence of the Friedmann's equations
Eq. (\ref{newton-friedman}) is that in exact general relativistic expansion dynamics of the universe there is no relativistic effects due to the velocity  of a receding galaxy. The expansion velocity is larger than the velocity of light for distances larger than the Hubble distance:
$V_{\mathrm{exp}}>c$ for $r>R_H$, where $R_H = c/H$ (see also Eq. (\ref{expvel})).

\subsection{The nature of cosmological redshift}
In the Sandage's list of the "23 astronomical problems"\cite{sandage95} the number fifteen (the first in the extragalactic section) sounds intriguingly:\textit{"Is the expansion real?"}.

In fact the literature on the SCM contains acute discussion on the nature of the cosmological redshift\cite{bartee12}, subject which constantly produces "common big bang misconceptions" or the "expanding confusions".  A summary of such discussions was done by Francis et al.\cite{francis07} who confronts Rees \& Weinberg claim: \textit{"how is it possible for space, which is utterly empty, to expand? How can nothing expand? The answer is: space does not expand. Cosmologists sometimes talk about expanding space, but they should know better"}, with the state by  Harrison\cite{harrison81}:\textit{ "expansion redshifts are produced by the expansion of space between bodies that are stationary in space"}.

In mathematical language within FLRW space expanding model the cosmological redshift is a new physical phenomenon where due to the expansion of space the wave stretching of the traveling photons occurs via the Lemaitre's equation
 $(1+z)=\lambda_0/\lambda_1 = S_0/S_1$,  which is different from the familiar in lab the Doppler's effect. One can also see this if compare relativistic Doppler and cosmological FLRW velocity-redshift V(z) relations. The relativistic Doppler relation has the form $V_{\mathrm{Dop}}(z) = c(2z+z^2)/(2+2z+z^2)$ and the velocity always less than $c$, while expanding space velocity $V_{\mathrm{exp}}$ can be arbitrary large\cite{bartee12}.

It is important to note that on the verge of modern technology there are direct observational tests of the physical nature of the cosmological redshift. First  crucial test of the reality of the space expansion was suggested by Sandage\cite{sandage62}, who noted that the observed redshift of a distant object (e.g. quasar) in expanding space must be changing with time according to relation
$dz/dt = (1+z)H_0 - H(z)$. In terms of radial velocity, the predicted change
$dv/dt \sim 1$ cm s$^{-1}$/yr. This may be within the reach of the future 42m ELT telescope\cite{liske08}.

Even within the Solar System there is a possibility to test the global expansion of the universe. According to recent papers by
Kopeikin\cite{kopeikin12,kopeikin14} the equations of light propagation used currently by Space Navigation Centers for fitting range and Doppler-tracking observations of celestial bodies contain some terms of the cosmological origin that are proportional to the Hubble constant $H_0$. Such  project as
PHARAO may be an excellent candidate for measuring the effect of the global cosmological expansion within Solar System, which
has a well-predicted frequency drift magnitude
$\Delta\nu /\nu =2H_0 \Delta t \approx 4\times 10^{-15} (H_0/70kms^{-1}Mpc^{-1})
(\Delta t/10^3s)$, where $H_0$ is the Hubble constant $\Delta t$ is the time of observations. In the case of the non-expanding Universe the frequency drift equals zero.

\subsection{Fractality of large-scale galaxy distribution}
Modern observations of the 3-dimensional galaxy distribution, obtained from huge redshift surveys (such as 2dF and SDSS),  demonstrate that at least for interval of  scales $1 \div 100$ Mpc/h there is a power law relation between the galaxy number density $n(R)$ and the radiuses of spheres $R$, so that $n(R) \propto R^{-\gamma}$ (see reviews by Sylos Labini\cite{sylos11} and
Baryshev \& Teerikorpi\cite{bartee12}). Such power law behavior is known as the
\textbf{de Vaucouleurs law}\cite{devac70}.
Note that the power law correlation function is the characteristic feature of the discrete stochastic fractal structures in physics (phase transitions, strange attractors, structure growth) and has clear mathematical presentation (e.g. Gabrielli et al.\cite{gabrielli05}).

At the same time modern observations of the \textbf{Hubble law} in the local Universe based on Cepheid distances to local galaxies, supernova distances, Tully-Fisher distances and other distance indicators, demonstrate that the perfect
linear Hubble law is well established within the same distance interval of scales $1 \div 100$ Mpc/h (e. g. Sandage\cite{sandage10},
Baryshev \& Teerikorpi\cite{bartee12}).

A puzzling conclusion is that the Hubble law, the strictly linear redshift-distance relation, is observed just inside strongly inhomogeneous galaxy distribution, i.e. deeply inside fractal structure at scales $1 \div 100$ Mpc/h.
This empirical fact presents a profound challenge to the standard model where the homogeneity is the basic explanation of the Hubble law, and \textit{"the connection between homogeneity and Hubble's law was the first success of the expanding world model"} (Peebles et al.\cite{peebles91}).

However, contrary to this expectation, modern data show a good linear Hubble law even for nearby galaxies. It leads to a new  conceptual puzzle that the linear Hubble law is not a consequence of the homogeneity. A cosmological model which can unite  the Hubble law with fractality of matter distribution and the cosmic microwave background radiation was suggested
by Baryshev\cite{bar08d}.

\section{Conclusions}
The positive role of the physical paradoxes in the science of the Universe is well known from the history of cosmology\cite{bartee12}.
Even the known phenomena may have different interpretations, each corresponding
to a specific choice of the basic framework able to explain key observations.
Theoretical physics is a developing subject and "new physics" may offer a variety of new cosmological applications. Finally, observations and theoretical understanding are always limited, hence even a quite credible world model has its limitations, too (in current cosmology  99.5\%  of the needed mass has unknown
nature).
These emphasize the importance of crucial observational tests as the only safe way to decide between alternative cosmological ideas.

\end{document}